\PassOptionsToPackage{numbers,sort&compress}{natbib}
\documentclass[sigconf,nonacm]{acmart}

\usepackage[utf8]{inputenc}
\usepackage[T1]{fontenc}
\usepackage{microtype}
\usepackage{booktabs}
\usepackage{amsmath}
\usepackage{xcolor}
\usepackage{url}

\setcopyright{none}
\hypersetup{
  hidelinks,
  pdftitle={High-Value Synthetic Supervision for Parameter-Efficient Adaptation of a Compact Japanese Speech Model},
  pdfauthor={Sidi Chang; Peiying Zhu}
}
\providecommand{\doi}[1]{\url{https://doi.org/#1}}

\newcommand{\metric}{\textsc{FCR}}

\begin{document}
\acmshorttitle{High-Value Synthetic Supervision}
\acmshortauthors{Chang \& Zhu}
\pagestyle{acmstyle}
\thispagestyle{firstpage}

\twocolumn[
\begin{@twocolumnfalse}
\begin{center}
{\LARGE\bfseries High-Value Synthetic Supervision for Parameter-Efficient Adaptation of a Compact Japanese Speech Model\par}
\vspace{1.0em}

\begin{tabular}{@{}c@{\hspace{2em}}c@{}}
{\large Sidi Chang\textsuperscript{*\,$\dagger$}} &
{\large Peiying Zhu\textsuperscript{*}} \tabularnewline
{\small schang@blossomai.co} & {\small peiying@blossomai.co} \tabularnewline
{\small Blossom AI Labs} & {\small Blossom AI} \tabularnewline
{\small Tokyo, Japan} & {\small San Francisco, CA, USA}
\end{tabular}
\vspace{1.2em}
\end{center}

\noindent\parbox{\textwidth}{
\small
\noindent\textbf{Abstract}\\[3pt]
Private domain speech is difficult to collect and redistribute, while compact models need task-specific supervision. We study an auditable synthetic pipeline that maps Japanese care handoffs directly to six-field structured notes. Using 182 synthetic training and development clips, we adapt a 1.47B audio model by full fine-tuning and rank-16 LoRA. On a 39-clip scenario-seed-disjoint synthetic test, an unadapted model obtains a model-judged factuality--recall score of 0.0500, full tuning 0.8664, and LoRA 0.8461. LoRA reaches 97.7\% of the full-tuning aggregate as a descriptive ratio while project telemetry reports 12.4M trainable parameters (about 0.85\% of the backbone). Both adaptations show large paired gains over the same base; the full-versus-LoRA interval crosses zero, and differing optimization settings preclude an equivalence claim. This is a parameter-efficient capability-acquisition result, not a device-performance result: latency, memory, energy, and real-time factor were not measured. All evaluation speech and targets are synthetic, references are model-proposed, and the judge is uncalibrated. The evidence shows that a compact model can acquire a narrow audio-to-structure transformation from a few hundred provenance-linked synthetic examples; it does not establish clinical validity, real-speech transfer, or superiority to a clean cloud system.
}

\vspace{1.5em}
\end{@twocolumnfalse}
]

\renewcommand{\thefootnote}{\fnsymbol{footnote}}
\footnotetext[1]{Both authors contributed equally to this research.}
\footnotetext[2]{Corresponding author.}
\renewcommand{\thefootnote}{\arabic{footnote}}
\setcounter{footnote}{0}

\section{Introduction}

Compact speech models are attractive when sensitive audio should remain local, yet specialization is constrained by scarce redistributable data. Japanese care handoffs are a demanding case: the input is compressed spoken language and the output must separate observations, assessment, actions, and plans. Prior work has studied speech recognition and information extraction for nursing shift changes \cite{suominen-etal-2015-handoff} and synthetic clinical speech for augmentation \cite{lindsay-etal-2022-synthetic,scroggins-etal-2025-synthetic}. We ask a narrower question: can a compact audio-native model acquire an end-to-end speech-to-structure capability from a few hundred designed synthetic examples?

Our unit of supervision is not only a waveform--label pair. Each item links a scenario seed, fact requirements, a Japanese spoken realization, rendered audio, a structured target, and quality metadata. Teacher-generated responses can transfer behavior into audio models \cite{held-etal-2025-diva}, while concise definitions can seed learning under weak supervision \cite{peng-etal-2024-incubating,li-etal-2024-xws}. These results motivate investing in the specification and audit trail around each example rather than assuming that parameter scale alone solves a narrow private task.

We compare the same 1.47B backbone unadapted, fully tuned, and adapted with rank-16 LoRA. Our contributions are: (i) a provenance-linked synthetic supervision object; (ii) a same-backbone, seed-disjoint capability-acquisition result with paired uncertainty; and (iii) a descriptive parameter-efficiency result with an explicit boundary around unmeasured device execution. The last point matters for on-device research: reducing trainable parameters can make adaptation more practical, but a local-capable architecture and an export path are not measurements of latency, memory, energy, or reliability.

\section{Synthetic supervision}

\paragraph{Task and aligned object.}
The input is a Japanese spoken handoff and the output is a strict JSON object with six strings: Focus, Subjective, Objective, Assessment, Intervention, and Plan. Each scenario carries a fact checklist specifying content expected in the note. Both the checklist and reference note are generated benchmark artifacts, not certified clinical gold. A record retains scenario identity and family, spoken transcript, audio, target, checklist, care setting, dialogue mode, an advisory ASR-consistency alert, and planted-defect provenance.

Planted omissions or fabrications are reserved for evaluation-workflow development and excluded from the fine-tuning manifests. The ASR alert is different: it is an uncalibrated consistency signal and remains in the data. Calling it an exclusion gate would be incorrect.

\paragraph{Frozen v5 data.}
The snapshot contains 221 faithful clips from 32 scenario seeds. Seed-group assignment gives 139 training clips from 20 seeds, 43 development clips from six seeds, and 39 held-out clips from six seeds. Training consumes the 182 train-plus-development clips. No seed crosses partitions, and the evaluated IDs exactly equal the held-out manifest. Fifty-two training/development clips and eight held-out clips carry the ASR alert.

The held-out set is seed-disjoint but not family-disjoint or balanced. It covers fall risk (13), meal intake (8), night wandering (14), and vital change (4), omitting two of six families. Dialogue modes are 30 two-party handoffs, six single-speaker reports, and three dictation prompts. These distributions bound generalization claims.

\paragraph{Existing input-realism evidence.}
A separate internal review subset contains 75 final-rubric judgments of spoken input and audio: 58 items were usable as-is, 17 usable with a minor issue, and none unusable; 72 were marked natural Japanese. There is one independent label per item and no inter-annotator agreement estimate. This evidence concerns selected synthetic inputs, not all training clips, reference correctness, or model outputs.

\section{Adaptation and evaluation}

\paragraph{Model variants.}
We use \texttt{LiquidAI/LFM2.5-Audio-1.5B-JP}, an approximately 1.47B Japanese audio/text model distributed under the LFM Open License v1.0 \cite{amini-etal-2025-lfm2,liquidai-2026-jp-model-card}. Audio is resampled to 16-kHz mono and generation is constrained to the six-field contract. Full tuning uses learning rate $2\times10^{-5}$, batch size 4, and 200 steps. LoRA uses rank 16, $\alpha=16$, dropout 0.2, learning rate $10^{-4}$, batch size 4, and 150 steps; adapters target the attention projections and \texttt{w1--w3}. Recovered wall times are 462.9 and 435.7 seconds. Hardware and RNG seeds were not preserved, so wall time is reported only as run provenance, not an efficiency comparison.

Project telemetry reports 12.4M trainable LoRA parameters, approximately 0.85\% of 1.47B. The count was not serialized in the recovered configuration and was not independently reconstructed from the checkpoint; we therefore treat it as project telemetry rather than a verified measurement.

\paragraph{Systems and metric.}
All systems are scored on the same 39 clips: the unadapted model, a transcription-plus-cloud structuring cascade, full tuning, and LoRA. A separate instruction model at temperature 0 returns factual correctness (FC), context recall (CR), hallucinations, and missed required facts. The frozen per-clip verdict export and its aggregation implementation confirm that FC and CR are separately elicited per-clip scalars and macro-averaged across the 39 clips, while Hall. and Missed sum the lengths of separately returned error lists. The count columns are therefore not denominators or weights for CR: a system can have fewer listed misses but lower macro-CR when the per-clip scores and variable-length lists differ. Raw judge text was not retained and the judge was not calibrated against a sufficiently large independently adjudicated clinical set. Counts are model-judge detections.

We compute $\metric=2\overline{\mathrm{FC}}\,\overline{\mathrm{CR}}/(\overline{\mathrm{FC}}+\overline{\mathrm{CR}})$. Thus FCR is the harmonic mean of the two macro means, not the mean of per-clip harmonic scores. Uncertainty uses a paired clip bootstrap: 20,000 resamples of 39 IDs, retaining paired system outputs and recomputing each aggregate harmonic score; percentile 95\% intervals use seed 20260806.

\paragraph{Baseline integrity.}
Twenty-one of 39 stored cascade outputs contain 143 Unicode replacement characters. Its score is 0.7981 on affected clips and 0.9006 on unaffected clips, a post-hoc association rather than a causal estimate. We keep the registered aggregate for traceability but do not use it to claim clean cloud parity.

\section{Results}

\begin{table}[t]
\centering
\small
\begin{tabular}{lrrrrr}
\toprule
System & FC & CR & \metric & Hall. & Missed \\
\midrule
Unadapted & 1.0000 & 0.0256 & 0.0500 & 0 & 184 \\
Cascade$^{\dagger}$ & 0.9128 & 0.7873 & 0.8454 & 36 & 97 \\
Full tuning & 0.9121 & 0.8251 & \textbf{0.8664} & 27 & 79 \\
LoRA & 0.9069 & 0.7929 & 0.8461 & 26 & 72 \\
\bottomrule
\end{tabular}
\caption{Results on the 39-clip synthetic held-out set. Counts are model-judge outputs. $^{\dagger}$Encoding-affected baseline.}
\label{tab:results}
\end{table}

The unadapted model receives perfect FC because it produces almost no substantive content; its CR of 0.0256 and 184 missed facts expose this abstention-like artifact. Full tuning reaches 0.8664 and LoRA 0.8461 (Table~\ref{tab:results}). Relative to the same base, paired gains are $+0.8164151$ (95\% CI $[0.7143885,0.8854931]$) and $+0.7961004$ ($[0.6841394,0.8774086]$). These intervals support capability acquisition on the frozen synthetic benchmark.

Full tuning exceeds LoRA by 0.0203147, but the paired interval $[-0.0293397,0.0729566]$ crosses zero. The parameter-efficiency result is descriptive: LoRA obtains 97.7\% of the full-tuning aggregate while project telemetry reports less than 1\% trainable parameters. This makes the adaptation axis relevant to resource-constrained model development, but it does not measure inference efficiency. The study is not a non-inferiority test, and the two adaptation settings differ in learning rate and step count, so we do not claim equivalence or a causal advantage from LoRA.

The nominal full-minus-cascade and LoRA-minus-cascade differences are $+0.0209988$ (CI $[-0.0271980,0.0698786]$) and $+0.0006841$ (CI $[-0.0648183,0.0655097]$). Both are inconclusive even before accounting for cascade corruption. The defensible statement is that adapted systems occupy a similar nominal range to this encoding-affected baseline.

\section{Scope and limitations}

The main empirical fact is a large same-backbone change after 182 synthetic clips. The experiment does not isolate a single causal pipeline component, but the supervision repeatedly aligns scenario semantics, spoken form, a constrained output schema, and fact coverage. Because there is no format-only ablation, the measured gain cannot be partitioned between learning the six-field JSON contract and learning semantic fact transfer. The supported data-centric claim is therefore modest: a compact model can learn the combined narrow transformation when examples explicitly represent it. The experiment does not show that synthetic data generally replaces model or dataset scale.

Several limitations prevent a stronger deployment claim. First, all training and evaluation speech is synthetic; no real-room, speaker, accent, or clinical transfer is tested. Second, model-proposed references and an uncalibrated judge define the benchmark contract. Third, the held-out set is small and incomplete, and the runs do not establish seed robustness. Fourth, the cascade is corrupted. Fifth, no device latency, peak memory, energy, power, or real-time factor is available. The model has a local-inference path, but this paper does not demonstrate efficient execution on any named device. A future device study should freeze hardware, precision, audio duration, warm-up, and decoding settings before measuring those quantities.

The corpus is wholly synthetic and contains no patient recordings or health-record data. The paper reports de-identified aggregate statistics from an existing lawful internal quality-review workflow and does not analyze reviewer identities or participant outcomes. Deployment would still require consent, security controls, retention limits, and professional verification of every draft. Synthetic scenarios can encode implausible practice and demographic omissions. Static benchmarks may also correlate weakly with interactive preferences \cite{li-etal-2025-mind}; the existing input-realism audit cannot substitute for output utility or safety evaluation. Audio, transcripts, targets, fact checklists, metadata, and provenance may be offered to qualified researchers under controlled access, a data-use agreement, security requirements, and applicable third-party terms. Academic and commercial access may use different licenses and fee schedules; no unrestricted public release is claimed.

\section{Conclusion}

A 1.47B Japanese audio model acquired a narrow speech-to-structure capability from 182 provenance-linked synthetic clips. Full and LoRA adaptation produced large paired gains over the same base, while LoRA retained 97.7\% of the full-tuning aggregate with a small reported trainable fraction. This is evidence for high-value synthetic supervision and parameter-efficient adaptation of a compact model. It is not evidence of clinical validity or measured on-device performance; those require real-speech, human-centered, and device-specific studies that are intentionally outside the present claim.

\bibliographystyle{plain}
\bibliography{references}
\end{document}